# Non-Conventional Thermal States of Interacting Bosonic Oligomers

Amichay Vardi[1,2], Alba Ramos[3,4,5], Tsampikos Kottos[5]
[1]Department of Chemistry, Ben-Gurion University of the Negev, Beer-Sheva 84105, Israel
[2]ITAMP, Harvard-Smithsonian Center for Astrophysics, Cambridge, Massachusetts 02138, USA
[3]Institute for Modeling and Innovative Technology, IMIT (CONICET - UNNE), Corrientes W3404AAS, Argentina
[4]Physics Department, Natural and Exact Science Faculty, National University of the Northeast, Corrientes W3404AAS, Argentina
[5]Wave Transport in Complex Systems Lab, Department of Physics, Wesleyan University, Middletown, CT-06459, USA

There has recently been a growing effort to understand in a comprehensive manner the physics and intricate dynamics of many-body and many-state (multimode) interacting bosonic systems. For instance, in photonics, nonlinear multimode fibers are nowadays intensely investigated due to their promise for ultra-high-bandwidth and high-power capabilities. Similar prospects are pursued in connection with magnon Bose-Einstein condensates, and ultra-cold atoms in periodic lattices for room-temperature quantum devices and quantum computation respectively. While it is practically impossible to monitor the phase space of such complex systems (classically or quantum mechanically), thermodynamics, has succeeded to predict their thermal state: the Rayleigh-Jeans (RJ) distribution for classical fields and the Bose-Einstein (BE) distribution for quantum systems. These distributions are monotonic and promote either the ground state or the most excited mode. Here, we demonstrate the possibility to advance the participation of other modes in the thermal state of *bosonic oligomers*. The resulting non-monotonic modal occupancies are described by a microcanonical treatment while they deviate drastically from the RJ/BE predictions of canonical and grand-canonical ensembles. Our results provide a paradigm of ensemble equivalence violation and can be used for designing the shape of thermal states.

Recent experimental and theoretical breakthroughs have initiated a renaissance of studies on interacting many-boson systems. Two of the most notable examples are complex nonlinear multimode photonic arrangements [1][2][3] and interacting ultracold atomic gasses in periodic lattice potentials [4][5][6][7]. These systems provide excellent testbeds for unravelling altogether new horizons and possibilities well beyond photonics and matter-wave frameworks. Moreover, they hold promise for novel technological developments. For instance, advancement in mode multiplexing/demultiplexing schemes allow substantial expansion of information capacity and power delivery capabilities of multimode fibers [2][8][9][10][11]. Likewise, the unprecedented control of atomic and molecular quantum gasses provides a promising platform for quantum simulations and the design of novel quantum information protocols [12][13][14].

While these emerging platforms could prove revolutionary, they inevitably pose a set of fundamental challenges. In case of classical fields, like electromagnetic radiation propagating in nonlinear multimode fibers, nonlinearities lead to multi-wave mixing processes through which the many modes can exchange energy via a multitude of possible pathways, often numbering in the trillions even in the presence of one hundred modes or so. Similarly, particle-particle interactions in quantum gasses, result in intractable many-body dynamics. Luckily, statistical mechanics and thermodynamics [15][16][17] provide a wealth of all-encompassing insights on the macroscopic properties of various phases of matter despite the sheer complexity involved in monitoring the individual dynamics of a large number of interacting particles in systems with many degrees of freedom.

One of the most fundamental assumptions of statistical mechanics is ensemble equivalence [15][16]. Regardless of the constraints imposed on the accessible states, i.e. of whether one considers microcanonical, canonical, grand-canonical, or any other statistical ensemble, the obtained statistical averages for any physical observable should be identical. Different ensembles differ only in their predictions about fluctuations [15][18][19][20][21]. In particular, when applied to the occupations of one-particle natural modes of many-boson systems, all formulations lead to the derivation of thermal states in the ubiquitous forms of a Rayleigh-Jeans (RJ) power distribution for classical fields or Bose-Einstein (BE) distribution for quantum particles. These monotonic distributions, which promote either the ground state at positive temperatures – a behavior consistent with the process of beam self-cleaning observed in experiments [22][23] – or the most excited state of systems with negative temperature, have been predicted in various multimode settings [17][21][24][25][26][27] and observed in numerous experiments [28][29][30][31][32].

Ensemble equivalence is guaranteed in the thermodynamic limit where the number of particles $N$ and the system's volume $V$ (number of degrees of freedom $M$) are increased indefinitely while keeping a fixed particle density $N/V$ ($N/M$). However, in the *mesoscopic* regime, the moderate number of degrees of freedom may lead to surprising differences between ensembles. In this respect, the above-mentioned photonic and cold atom setups enable an *in vivo* study of the potential break-down of this fundamental concept. Since these systems can be typically considered closed and isolated over very long times, the total many-body energy and the total number of particles (or total light intensity in photonic realizations) can be viewed as motional constants, thereby singling out the microcanonical description over all other statistical ensembles. To the extent that its dynamics is sufficiently ergodic/chaotic, any initial many-body preparation will eventually spread uniformly throughout its respective energy shell at fixed $N$.

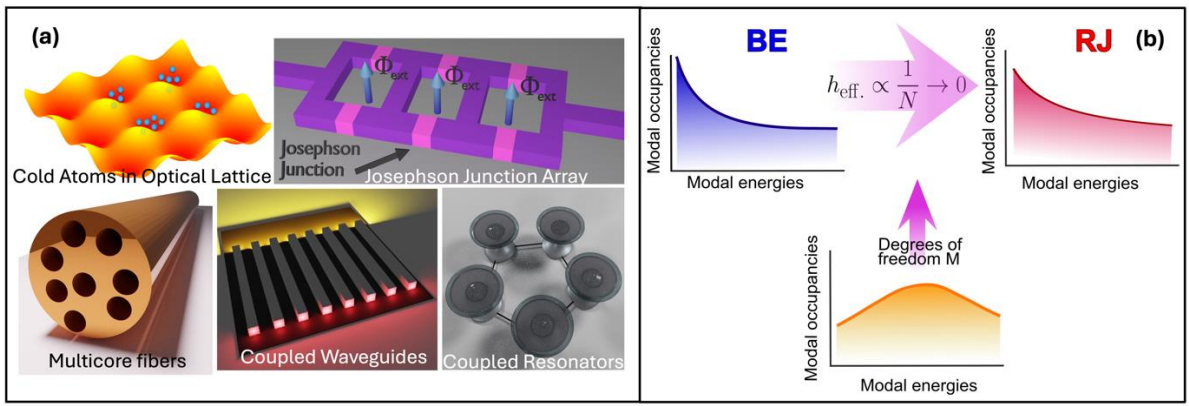

Figure 1: (a) Examples of bosonic physical systems that are described by a BHH (upper row) and its classical analogue (lower row). (b) The microcanonical thermal state of an oligomer (moderate number $M$ of degrees of freedom) can be described by a non-monotonic occupation statistics. As the thermodynamic limit is approached by increasing $M$, the thermal state turns to either a BE (when the number of particles/modes $N$ is small) or to a RJ (for $N \to \infty$).

Here we consider mesoscopic many-body bosonic oligomers with a moderate number of degrees of freedom $M$, and study their thermal state. The physical constraints of the model are microcanonical. Correspondingly, both classical mean-field simulations and quantum many-body time evolution result in long time occupations of the one-particle modes, that agree with a microcanonical distribution over many-body states, with no memory of the initial preparation.

However, this microcanonical occupation statistics differs dramatically from the RJ or BE distributions obtained for classical multimode photonic circuits and quantum many-body systems respectively, in the canonical as well as in the grand-canonical treatments. In particular, we find an energy range in which microcanonical thermalization results in *non-monotonic* occupation statistics, thereby promoting other modes than the ground state or the most-excited state. This anomalous domain vanishes in the thermodynamic limit as the number of freedoms increases and the microcanonical occupation statistics coincides with RJ or BE. Our results are easily implemented using current photonic circuits and cold-atom optical lattices and can be used for the design and manipulation of different types of thermal states via structure engineering.

**Results**
*Theoretical Modeling of Quantum and Classical Bosonic Oligomers* – The simplest non-trivial description of $N$ interacting bosons confined in a lattice of $M$ sites is the tight binding Bose-Hubbard Hamiltonian (BHH) [33]. Our analysis here focuses on this model due to the broad variety of its applications ranging from quantum optics and BEC optical lattices (OL) to Josephson arrays and cantilever vibrations in micromechanical arrays, see Fig. 1. In second quantization it reads:

$$\hat{H} = -\sum_{l,m=1}^{M} \hat{b}_l^\dagger J_{lm} \hat{b}_m + \frac{1}{2} U \sum_l \hat{n}_l (\hat{n}_l - 1); \quad \hbar = 1 \quad (1)$$

where $l, m = 1, \cdots, M$, is the number of nodes (sites), $J_{lm} = J_{ml}^*$ is the connectivity matrix that dictates the couplings between sites $l$ and $m$, $U$ is the particle-particle interaction strength, $\hat{n}_l = \hat{b}_l^\dagger \hat{b}_l$ is the boson number operator at site $l$ and $\hat{b}_l^\dagger, \hat{b}_l$ obey canonical commutation rules $[\hat{b}_l, \hat{b}_m^\dagger] = \delta_{lm}$. Below we introduced a weak randomness on the on-site potential $J_{l,l}$ to avoid degeneracies, while $J_{l \neq m} = \delta_{l,m\pm 1}$. In the supplementary material (SM) we report results with other topologies [44].

Interacting bosonic systems described by the BHH Eq. (1) have a well-defined classical limit. By rescaling the lowering and raising operators $\hat{\psi}_l = 1/\sqrt{N} \hat{b}_l$ and $\hat{\psi}_l^\dagger = 1/\sqrt{N} \hat{b}_l^\dagger$, the commutation relation takes the form $[\hat{\psi}_l, \hat{\psi}_m^\dagger] = (1/N)\delta_{lm}$ where $\hbar_{eff} = 1/N$ plays the role of an effective-$\hbar$. The classical limit $\hbar_{eff} \to 0$ is then identified as $N \to \infty$ while keeping $\chi = U \cdot N$ constant. In this case, the $\hat{\psi}_l$ operators become $c$-numbers, resulting in a classical Hamiltonian $\mathcal{H}$

$$\mathcal{H} \equiv \frac{H}{N} = -\sum_{m \neq l} \psi_l^* J_{lm} \psi_m + \frac{1}{2} \chi \sum_l |\psi_l|^4 ; \text{where} \sum_l |\psi_l|^2 = 1 \quad (2)$$

The classical Hamilton's equations of motion for the Hamiltonian in Eq. (2) read,

$$i \frac{\partial \psi_l(t)}{\partial t} = -\sum_{m \neq l} J_{lm} \psi_m(t) + \chi |\psi_l(t)|^2 \psi_l(t). \quad (3)$$

These are used to model the field dynamics of nonlinear photonic networks [34][35][36]. Examples of such systems include, coupled micro-resonators, coupled waveguides, multicore/multimode fibers etc, see Fig. 1. For the last two examples, the time variable $t$ plays the role of the paraxial direction $z$. In these optical realizations, $\psi_l$ describes the complex field amplitude at node $l$ while the last term in Eq. (3) describes the nonlinear light-matter interactions due to a Kerr effect.

*Orbital basis representation* – Equations (1,2) describe the systems in the local (Wannier) basis associated with the (localized) modes of the individual nodes of the system. Transformation into the natural modes of the underlying non-interacting network is obtained by diagonalization of the connectivity matrix $J$, resulting in the eigenvectors $\{f_\alpha\}, \alpha = 1, \cdots, M$ associated with the natural mode energies $\varepsilon_\alpha$. Below, we refer to this basis as the *orbital basis* or the *supermode basis* and denote the $l$-th component of the $\alpha$-th eigenvector by $f_\alpha(l)$. The corresponding field operators in the orbital basis are $\hat{a}_\alpha = \sum_l f_\alpha(l)\, \hat{b}_l$ resulting in an orbital number operator,

$$\hat{v}_\alpha = \hat{a}_\alpha^\dagger \hat{a}_\alpha = \sum_{l,m} f_\alpha^*(l) f_\alpha(m) \hat{b}_l^\dagger \hat{b}_m, \quad \alpha = 1, \dots, M. \quad (4)$$

The BHH Eq. (1) in the orbital basis then reads,

$$\hat{H} = \sum_{\alpha=1}^{M} \varepsilon_\alpha \hat{a}_\alpha^\dagger \hat{a}_\alpha + \frac{U}{2} \sum_{\alpha\beta\gamma\delta} \Gamma_{\alpha\beta\gamma\delta} \hat{a}_\alpha^\dagger \hat{a}_\beta^\dagger \hat{a}_\gamma \hat{a}_\delta; \quad \hbar = 1, \quad (5)$$

where $\Gamma_{\alpha\beta\gamma\delta} = \sum_{\alpha=1}^{M} f_\alpha^*(l) f_\beta^*(l) f_\gamma(l) f_\delta(l)$ describes the interactions associated with the nonlinear four-wave mixing between supermodes. Similarly, the classical Hamiltonian Eq. (2) is transformed by $\psi_l(t) = \sum_l f_\alpha(l)\, C_\alpha(t)$ into,

$$\mathcal{H} = \sum_{m \neq l} \varepsilon_\alpha |C_\alpha|^2 + \frac{\chi}{2} \sum_l \Gamma_{\alpha\beta\gamma\delta} C_\alpha^* C_\beta^* C_\gamma C_\delta. \quad (6)$$

*Constants of motion and thermal states in the thermodynamic limit* – We study the classical dynamics of the model Hamiltonian of Eqs. (2,6) as well as the full quantum dynamics of Eqs. (1,5). The system is launched in an arbitrary non-equilibrium state and we are interested in the long time orbital occupation statistics. Both classical and quantum motion conserve the total energy (longitudinal electromagnetic momentum flow in fiber/waveguide settings), denoted respectively as $\mathcal{H}(\{\psi_l(t)\}) = E$ and $\langle\hat{H}\rangle_{\psi(t)} = \langle\psi(t)|\hat{H}|\psi(t)\rangle = E$, as well as the total number (total optical power of a beam in the classical optics framework) $\mathcal{N}(\{\psi_l\}) = \sum_l |\psi_l|^2 = P = \sum_\alpha |C_\alpha|^2$ and $\langle\psi(t)|\hat{N}|\psi(t)\rangle = N$, where $|C_\alpha|^2$ is the normalized optical power in the orbital $\alpha$ and $\hat{N} = \sum_{l=1}^{M} \hat{b}_l^\dagger \hat{b}_l = \sum_{\alpha=1}^{M} \hat{v}_\alpha^\dagger \hat{v}_\alpha$ is the number operator. We set the normalization to $P = 1$.

In the spirit of the thermodynamic paradigm, we limit our calculations to the regime where the interaction/non-linear part of the Hamiltonian is small with respect the linear part so that $\mathcal{H}(\{\psi_l(t)\} = \mathcal{H}_L + \mathcal{H}_{NL} = E \approx \mathcal{H}_L = \sum_\alpha \varepsilon_\alpha |C_\alpha|^2$. To the extent that the interaction term is still sufficient to generate chaos, the system dynamics will be ergodic. One classical trajectory would ergodically explore the energy shell $E \pm \delta E$ of the non-interacting system, whereas a semiclassical cloud of trajectories or a quantum probability distribution would spread out throughout this shell. In the limit where the number of modes is large, the resulting mean orbital occupations in quantum- and classical simulations should follow the Bose-Einstein (BE) and Rayleigh-Jeans (RJ) statistics respectively [17][21][24][25][26][27][28][29][30][31][32]

$$\bar{v}_\alpha = \frac{1}{e^{(\varepsilon_\alpha - \mu)/T} - 1} \text{ (BE)}; \quad \langle|C_\alpha|^2\rangle = \frac{T}{\varepsilon_\alpha - \mu} = \bar{v}_\alpha, \text{ (RJ)} \quad (7)$$

where the temperature $T$ and chemical potential $\mu$ are set by the two conserved quantities i.e. the total internal energy and total optical power/number of particles.

The BE and RJ distributions can also be derived using a canonical or a grand-canonical formalism, however without requiring explicitly large $M$. Both distributions are monotonic functions of the orbital modes and either promote the ground state (positive temperature) or the most excited state (negative temperature). In fact, the promotion of the ground state in case of positive temperatures has been nicely confirmed in optics experiments with multimode nonlinear fibers where the phenomenon of beam self-cleaning was observed [22][23]. Our goal here is to show how the moderate number of degrees of freedom in a mesoscopic oligomer system may produce microcanonical mean orbital-occupations that deviate from the RJ and BE prediction, thus violating ensemble equivalence. Importantly, as shown below, we can use $M$ and $E$ as engineering knobs for the design of non-monotonic mean occupations that promote the participation of specific orbitals in the thermal state.

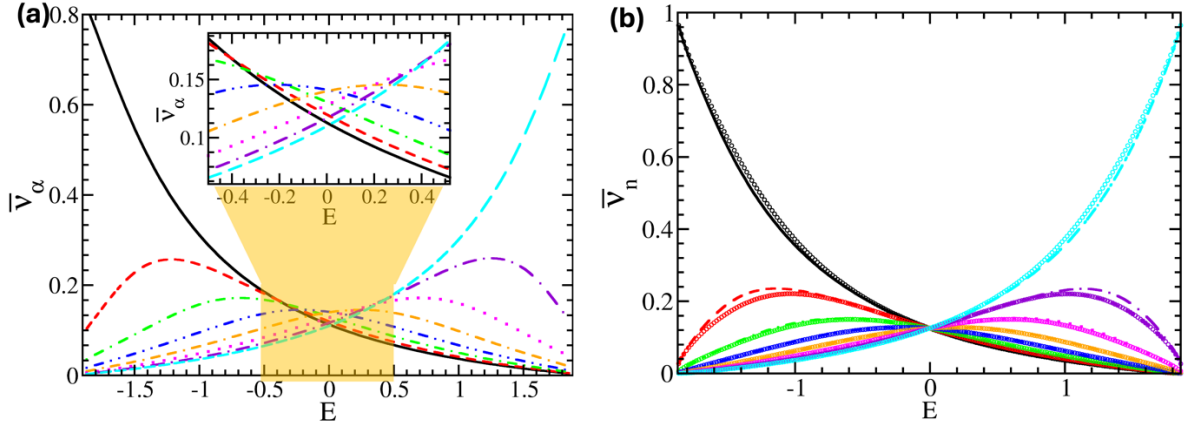

Figure 2: (a) Microcanonical predictions for the modal occupations $\bar{\nu}_\alpha$ versus energy for an oligomer with $M = 8$ and $N = 50$. Various lines indicate the various modes: $\bar{\nu}_1$(solid black), $\bar{\nu}_2$(dashed red), $\bar{\nu}_3$(dashed-dotted green), $\bar{\nu}_4$(dashed- double dotted blue), $\bar{\nu}_5$(doubled dashed- dotted orange), $\bar{\nu}_6$(dotted magenta), $\bar{\nu}_7$(long dashed-dotted violet), $\bar{\nu}_1$(long-dashed light blue), The orange highlighted domain indicates the energy range where the scrambling of the modal occupancies occur. In this domain the microcanonical modal occupancies differ drastically from the BE or the RJ statistics (Inset) Magnification of the scrambling domain. (b) The same as in (a) but now utilizing a canonical ensemble. The canonical treatment does not reveal the formation of a scrambling regime. Instead, it follows the BE predictions (colored circles).

*Microcanonical Thermal States of Bosonic Oligomers* – Our model explicitly conserves the total number of particles/optical-power and total internal energy. Hence it is appropriate to use a microcanonical formalism for the calculation of orbital occupancies $\{\bar{\nu}_\alpha\}$. The total energy of the non-interacting system is $E = \sum_{\alpha=1}^{M} \varepsilon_\alpha \nu_\alpha$ where $\varepsilon_\alpha$ are the orbital (optical mode) energies (propagation constants) and $\nu_\alpha$ are the orbital occupations which may be discretized as positive integers $\nu_\alpha \in Z^+$, subject to the constraint $\sum_{\alpha=1}^{M} \nu_\alpha = N$, for a total of $D = \frac{(N+M-1)!}{N!(M-1)!}$ combinations. In the quantum context, these integer occupations $r = (\nu_1, \nu_2, \cdots, \nu_M)$ are the good quantum numbers denoting the orbital Fock basis $|r\rangle = |\nu_1, \nu_2, \cdots, \nu_M\rangle$, i.e. the energy eigenbasis of the interaction-free Hamiltonian $\hat{H}_L = \sum_{\alpha=1}^{M} \varepsilon_\alpha \hat{a}_\alpha^\dagger \hat{a}_\alpha$, and $D$ is the Hilbert space dimension. The microcanonical occupation distribution $\bar{\nu}_\alpha$ is obtained by averaging the mode occupations over all such configurations that lie within the energy shell $E_r = \sum_{\alpha=1}^{M} \varepsilon_\alpha \nu_\alpha \in [E_0 + \delta E, E_0 - \delta E]$, with $\delta E$ being the shell-width.

In Fig. 2a the resulting microcanonical occupation statistics is shown for an oligomer with $M = 8$ sites. The mean occupation numbers are plotted versus the central shell energy $E$. We find that there is a parametrically large energy domain $\Delta_s \equiv E_s^{max} - E_s^{min}$ where $\{\bar{\nu}_\alpha\}$'s are scrambled i.e. energetically higher orbital modes in the middle of the spectrum have higher occupations than the ground state (or, in energy regions with negative temperatures, the most excited state). This behavior must be contrasted with the well-known canonical and grand-canonical pictures which predict BE/RJ monotonic occupation distribution. For the grand-canonical ensemble that assumes known *means* of the total energy and particle/power, the grand-partition function factorizes into a product of mode functions that take the BE form, or in the continuum limit, the RJ form, *regardless*

*of the number of* modes. Similar derivation exists for the canonical ensemble that replaces energy conservation by a known mean energy but strictly conserves particle number.

In Fig. 2b we plot the canonical occupation distribution $\bar{v}_\alpha = \frac{\sum_r v_\alpha(r) e^{-\beta E_r}}{\sum_r e^{-\beta E_r}}$ where $v_\alpha(r)$ denotes the occupation of the $\alpha$-th orbital in the $r$-th many-body state, obtained for an oligomer with $M = 8$ using the same discretized basis used for the microcanonical calculation. The parameter $\beta$ tunes the mean energy $\bar{E} = \frac{\sum_r E_r e^{-\beta E_r}}{\sum_r e^{-\beta E_r}}$ to be equal to $E$. It is clear that the canonical distribution does not maintain the non-monotonic nature predicted by the microcanonical ensemble. Instead, it retrieves the BE/RJ form also predicted by the grand-canonical analysis, where the distributions are monotonic functions of the modal energies. The substantial deviation from the microcanonical calculation demonstrates that the observed violation of the ensemble equivalence in case of oligomers and the transition from monotonic to non-monotonic means comes from the canonical relaxation of the energy constraint rather than from the grand-canonical relaxation of the number of particles constraint.

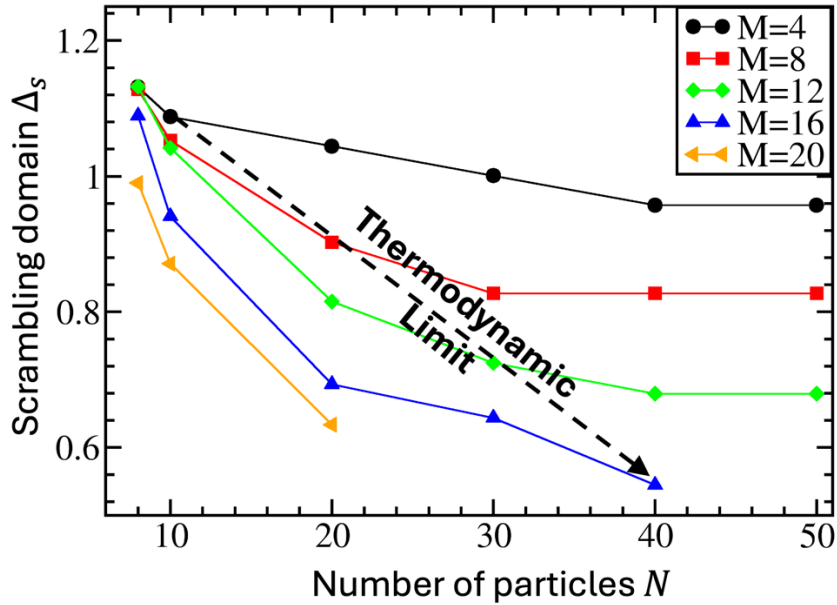

Figure 3: The scrambling domain $\Delta_s$ versus the number of particles $N$. For moderate degrees of freedom $M$ the scrambling domain reach a saturation value which is different than zero. At the thermodynamic limit $M \to \infty$ *while* $N/M$=const. (indicated with a dashed line) the scrambling domain becomes zero $\Delta_s = 0$ and the occupation statistics turns to a BE or to a RJ.

The disagreement between the microcanonical and canonical occupation means becomes clear when one considers the canonical partition function $Z = \sum_r e^{-\beta E_r} = \sum_E \Omega(E) e^{-\beta E}$. In the thermodynamic limit, the explosive increase of the density of states $\Omega$ with $E$ and the exponential decrease of the Boltzmann factor $e^{-\beta E}$ (or the decrease of the former and the increase of the latter in negative temperature regimes) combine to give a narrow integrand that only has marginal tails outside the microcanonical shell determined by the choice of $\beta$. However, with a mesoscopic number of degrees of freedom, the increase of the density of states is much more moderate, giving far longer tails that skew the microcanonical averages. This is particularly true in low $\beta$, high (positive or negative) temperature regimes, such as the central energy section of Fig. 2 where the deviations are observed. For more details on the entropy and temperature of the Bose Hubbard oligomer system, see SM [44].

Ensemble equivalence is of course re-established in the thermodynamic limit, where all statistical treatments including the microcanonical one, result in BE/RJ. In Fig. 3, we analyze the dependence of the parametric energy range with respect to the number of sites $M$ (optical volume) and number of particles $N$ (optical power). We find that as $\hbar_{eff} \to 0$ ($N \to \infty$), the energy domain $\Delta_s$ where scrambling is observed, reaches a saturation value. This saturation value goes to zero $\Delta_s \to 0$ in the thermodynamic limit. This behavior of bosonic oligomers is generic and independent on the underlying connectivity (dimensionality) of the network. For example, we have confirmed that this non-conventional behavior occurs also in cases where the connectivity matrix is described by full Random Matrix Theory (see SM [44]).

*Quantum Mechanical Simulations of Thermalization* – Having established the form of the thermal occupation distribution, we proceed to investigate whether they will actually be obtained if we prepare the system in a non-equilibrium state and let it evolve in time. For the quantum evolution we consider the case where the system is launched in any of the interaction-free eigenstates $|r_0\rangle$, hence the dynamics in the presence of interaction is restricted to fixed $N = \langle r_0|\widehat{N}|r_0\rangle$ and a narrow energy shell of width $\delta E \approx \sqrt{\langle r_0|\widehat{H}^2|r_0\rangle - \langle r_0|\widehat{H}|r_0\rangle^2}$ around $E = E_{r_0}$. We have verified that in all simulations the interaction can be treated as a small perturbation, i.e. $\delta E \ll E$.

Denoting the exact eigenstates and eigenvalues of the total Hamiltonian $\widehat{H}$ as $|m\rangle$ and $\mathcal{E}_m$ respectively, we evaluate the probability to find the system in any of the unperturbed eigenstates $|r\rangle$ at a later time $t$ as:
$$P_{r,r_0}(t) = \sum_{m,m'} \langle m'|r\rangle\langle r|m\rangle\langle m|r_0\rangle\langle r_0|m'\rangle e^{-i(\mathcal{E}_m - \mathcal{E}_{m'})t}, \quad (8)$$
The oscillating terms cancel out after sufficiently long time resulting in:
$$P_{r,r_0}(\infty) = \sum_m \mathcal{L}_{r,m} \mathcal{L}_{r_0,m}, \quad (9)$$
where $\mathcal{L}_{r,m} = |\langle m|r\rangle|^2$ is the local density of states (LDOS), i.e. the projection of the unperturbed eigenstate $|r\rangle$ onto the exact eigenstates $|m\rangle$. The one-particle occupations are then,
$$\langle \nu_\alpha \rangle_t = \sum_{r=1}^D P_{r,r_0}(t) \nu_\alpha(r) \to \bar{\nu}_\alpha = \sum_{r=1}^D P_{r,r_0}(\infty) \nu_\alpha(r). \quad (10)$$

The long-time occupation distribution is thus completely determined by the LDOS of the initial state, which we obtain by direct diagonalization of the BHH oligomers. Results were obtained for $M$ in the range of four to eight sites. Due to the poor scaling of the Hilbert space dimension $D$ with $N$, maximal particle numbers were restricted from about 50 particles with $M = 4$ to 10 particles with $M = 8$. Chaoticity was determined independently by energy-resolved analysis of the level spacing statistics [37] and by the participation number $PN_{r_0} = \left(\sum_m \mathcal{L}_{r_0,m}^2\right)^{-1}$ of the initial state $|r_0\rangle$ in the exact eigenbasis $|m\rangle$ [38][39][40] (see SM [44]). The latter quantity is predicted by random matrix theory to attain the value $PN_{GOE} = D_s/3$ for a Gaussian Orthogonal Ensemble (GOE) [41], where $D_s$ denotes the Hilbert space dimension of the pertinent energy shell.

Representative results are plotted in Fig. 4 for an oligomer with $M = 8$ and $N = 10$. Each set of eight mode occupations corresponds to the final distribution obtained after starting from one of the $D = 19448$ unperturbed eigenstates, using the prescription of Eq. (10). It is evident that all initial states in shells that lie in the chaotic (high *PN*) domain approach the same final occupation distribution, regardless of the details of the initial preparation. This small state-to-state variance of the occupations is a clear indication of thermalization. The resulting thermal distribution deviates significantly from the BE expectation (blue dashed lines) and agrees well with the predictions of the microcanonical ensemble (green diamonds), thus confirming the appearance of a scrambling domain $\Delta_s$ for total energies $E$ where thermalization is expected. For energies outside the chaotic domain, ergodization is lost and the final mode-occupation outcome depends on the initial preparation, i.e. the system does not thermalize due to the appearance of quasi-integrable regions within the relevant energy shells.

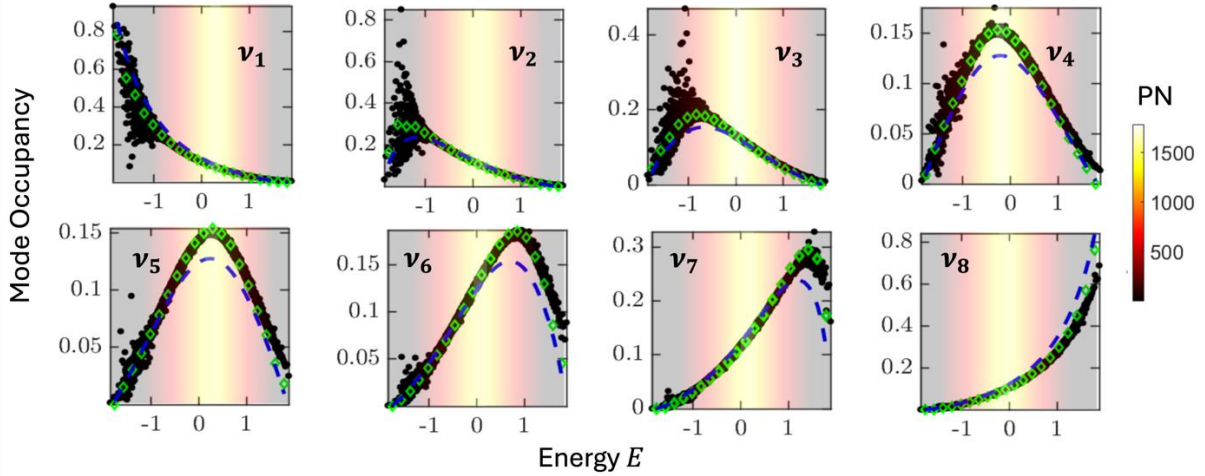

Figure 4: Modal occupations $v_\alpha$ versus the energy shell $E$ for a BHH with $M = 8, N = 10$ and $\chi = 3$. Quantum calculations (black filled circles), microcanonical treatment (green diamonds) and BE statistics (blue dashed line). The "heat map" indicates atypical value of the $PN_v$ of the corresponding eigenmodes. The regimes of highest $PN_v$ (yellow highlighted domains) indicate an ergodic spreading of the unperturbed states while a low $PN_v$ values (black highlighted areas) indicate absence of ergodic spreading. In the former case the system can reach thermalization while in the latter it does not. The deviation of the quantum calculations from the BE expectations is apparent for $v_{3,4,5,6}$ in the scrambling energy range $E \approx [-1,1]$. Instead, the quantum results agree nicely with the microcanonical calculations in the scrambling domain $\Delta_s$.

*Classical Simulations* – We stress that the non-conventional form of the occupation statistics is a mesoscopic effect due to the small number of degrees of freedom rather than a quantum effect due to the finite $N$ discretization. In particular, the same non-monotonic distributions are obtained in the purely classical dynamics generated by the Hamiltonian of Eq. (2). Consider, for example, the thermal power distribution of an initial beam excitation to the supermodes of a weakly nonlinear multi-mode/core fiber. All three ensembles predict a RJ in the thermodynamic limit corresponding to large number of modes $M$. Such RJ distribution has been confirmed experimentally by a number of groups [28][29][30][31].

We investigated the classical thermal occupation statistics in case of oligomers by integrating numerically Eq. (3) for long times, see SM [44]. The supermode amplitudes $C_\alpha(t)$ have been evaluated from the projection of $\psi_l(t)$ on the supermode basis $\{f_\alpha\}$. The thermal occupations $\langle |C_\alpha|^2 \rangle$ have been evaluated by calculating a time-averaged value of the modal occupations $\langle |C_\alpha|^2 \rangle_t$, after an initial transient time $t_{min}$ has been lapsed. Additionally, we have performed an average over 200 initial preparations $\{\psi_l(t=0), l = 1, \cdots, M\}$ that satisfy the energy and normalization constraints that define the specific state. A clear convergence to a corresponding steady state value is signaling that the system has reached its equilibrium state.

The dynamical results for $\langle |C_\alpha|^2 \rangle$ are compared in Fig. 5 to the theoretical predictions of RJ Eq. (7) and to the results from the microcanonical ensemble (Fig. 2a). Each panel in this plot corresponds to presenting the occupations at a single $E$ value in Fig. 2 and Fig. 4, where the chosen energy values all lie in the scrambling region. In all cases, we have confirmed that the dynamics is chaotic by evaluating the maximum Lyapunov exponent, which has been found to be positive (see SM [44]). The dynamical results for the modal occupancies and the microcanonical predictions show an enhanced participation of an intermediate orbital mode in the thermal state. This has to be contrasted with the canonical and grand-canonical expectations that prescribe a monotonic mean occupation distribution. As expected, the disagreement is more pronounced in the $\beta \approx 0$ central part of the scrambling domain (corresponding to total energy in the middle of the spectrum $E \approx 0$, see SM [44]) where the non-conventional equilibrium state is dominated by modes that are in the middle of the orbital spectrum.

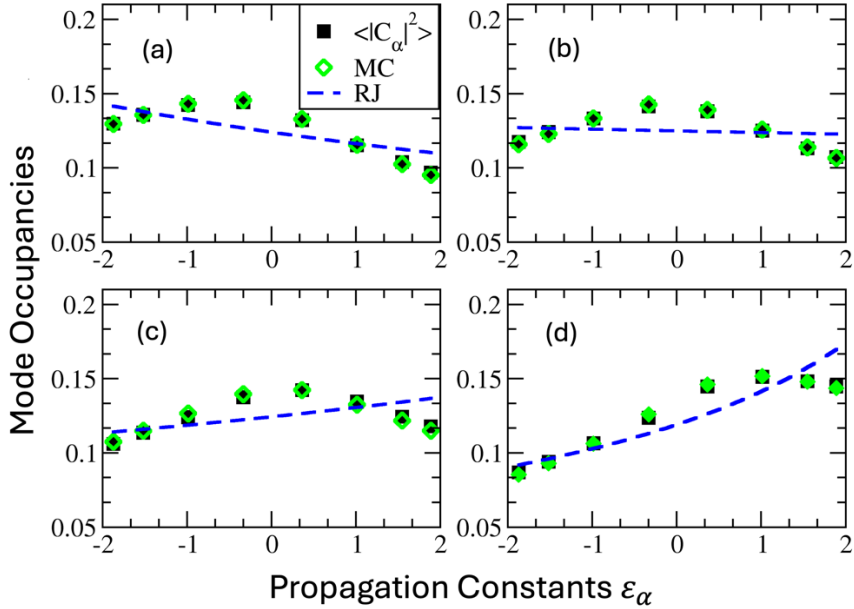

Figure 5: Modal power occupancies at thermal equilibrium of an initial beam excitation with total initial power $P = 1$ and various internal energy values: (a) $E \approx -0.103$ ; (b) $E = -0.03$ ; (c) $E = 0.097$ ; (d) $E = 0.297$. The results of the dynamical simulations are indicated with black square, the microcanonical treatment is indicated with green diamonds and the RJ statistics with blue dashed lines. As the total internal energy increases towards the center of the scrambling domain, the simulations show a stronger deviation from the RJ predictions. Instead, the microcanonical calculations describe nicely the numerical data in all energy regimes. The system consists of $M = 8$ modes with nonlinear strength $\chi = 3$.

While our analysis is focused on systems with nearest neighbor couplings between sites, similar non-conventional thermal distributions are found for other configurations of oligomers. For example, in the SM [44] we also present the quantum and classical simulations of fully connected oligomers which also demonstrate strong deviations from the BE/RJ predictions. In all cases, the microcanonical treatment accurately captures the dynamical results, thus demonstrating the violation of the fundamental principle of ensemble equivalence for the description of mean modal occupancies in case of mesoscopic systems with few degrees of freedom.

*Conclusions* – We have demonstrated the existence of non-conventional thermal states with non-monotonic occupation statistics occurring in interacting bosonic oligomers. This anomalous thermal state is pronounced for energies in the middle of the spectrum and is captured by microcanonical considerations. The latter agrees with both quantum and classical simulations. The energy domain where the non-monotonic occupation statistics occurs contracts to zero in the thermodynamic limit where the traditional BE (for quantum systems) or RJ (for classical field) is recovered. On the fundamental level, our results set constraints on the validity of ensemble equivalence in case of systems with few degrees of freedom. On the practical level the establishment of a non-conventional thermal state opens up new avenues for designing high-power optical fibers and atomic lattices that can result in thermal distributions that promote occupations of modes different from the ground or most excited ones.

# Supplementary Material

Non-Conventional Thermal States of Interacting Bosonic Oligomers

Amichay Vardi[1,2], Alba Ramos[3,4,5], Tsampikos Kottos[5]

[1]Department of Chemistry, Ben-Gurion University of the Negev, Beer-Sheva 84105, Israel
[2]ITAMP, Harvard-Smithsonian Center for Astrophysics, Cambridge, Massachusetts 02138, USA
[3]Institute for Modeling and Innovative Technology, IMIT (CONICET - UNNE), Corrientes W3404AAS, Argentina
[4]Physics Department, Natural and Exact Science Faculty, National University of the Northeast, Corrientes W3404AAS, Argentina
[5]Wave Transport in Complex Systems Lab, Department of Physics, Wesleyan University, Middletown, CT-06459, USA

*Time Dependent Coupled Mode Equation* – Using the orthogonal transformation $\psi_l(t) = \sum_l f_\alpha(l) C_\alpha(t)$ we can re-write the time-dependent CMT, Eq. (3) of the main text, in the orbital basis as

$$i\frac{\partial C_\alpha(t)}{\partial t} = \varepsilon_\alpha C_\alpha(t) + \chi \sum_{\beta\gamma\delta} \Gamma_{\alpha\beta\gamma\delta} C_\beta^*(t) C_\gamma(t) C_\delta(t) \quad (SM1)$$

where $\Gamma_{\alpha\beta\gamma\delta} = \sum_{\alpha=1}^M f_\alpha^*(l) f_\beta^*(l) f_\gamma(l) f_\delta(l)$.

While one can directly integrate Eq. (SM1) to extract the $C_\alpha(t)$ ($\alpha = 1, \cdots, M$), we have instead integrated numerically Eq. (3) and then projected the $\{\psi_l(t)\}$ in the supermode basis $\{f_\alpha\}$. The numerical integration was performed using a high-order three-part split symplectic integrator scheme (see Ref. [21] of the main text). The method conserved, up to errors $\mathcal{O}(10^{-8})$ the total energy $E$ and optical power $P$ of the system. These quantities have been monitored during the simulations to ensure the accuracy of our results. In all our simulations we have confirmed that the nonlinear contribution to the total energy $\mathcal{H}_{NL} = \frac{\chi}{2}\sum_l |\psi_l|^4 \ll \mathcal{H}_L = -\sum_{m \neq l} \psi_l^* J_{lm} \psi_m$.

In practice, the approach to a thermal equilibrium state involves a long-time propagation of an initial preparation $\{\psi_l\}$ ($l = 1, \cdots, M$). Typical integration times were as long as $10^8$ coupling constants. After an initial transient time $t_{min} \sim 5 \times 10^7$, we have calculated a time-averaged value of the modal occupations

$$\langle |C_\alpha|^2 \rangle_t = \frac{1}{t - t_{min}} \int_{t_{min}}^t |C_\alpha(t)|^2 dt. \quad (SM2)$$

We have also performed an additional average over 200 different initial configurations $\{\psi_l(t=0)\}$ chosen in a way that satisfy the total energy and power constraints $E, P$ respectively. A convergence of the modal occupation $\langle |C_\alpha|^2 \rangle_t$ to a steady state value has been considered as an (indirect) indication that the system has reached a thermal equilibrium state. Finally, in all our simulations we have used the normalization condition $P = \sum_l |\psi_l|^2 = 1$ for the total power.

*The oligomer spectrum, microcanonical entropy and temperature* – The interaction free spectrum $\{E_r\}$ and the exact spectrum $\{E_m\}$ of an oligomer with $M = 8$, $N = 10$ (total Hilbert space dimension $D = 19448$) and $\chi = 3.0$ are shown in Fig. SM1(a) along with the interaction energies $E_m^{Int} = \langle m|\hat{H}_{Int}|m\rangle$ and $E_r^{Int} = \langle r|\hat{H}_{Int}|r\rangle$. We find that the contribution of interactions to the total energy is small, justifying the perturbative picture in which interactions only serve to induce dynamics on the unperturbed energy shell. The mean spacing between adjacent levels $\bar{s}$ is plotted in Fig. SM1(b) and the corresponding microcanonical entropy $S_{mc} = k_B \ln \Omega$ where $\Omega = 1/\bar{s}$ is the density of states, is shown in Fig. SM1(c). The resulting inverse temperature parameter $\beta = \partial \ln \Omega / \partial E$ [Fig. SM1(d)] exhibits a mid-spectrum transition from positive to negative temperature, typical to spin-like systems with bound spectra (see Ref. [42] of the main text). The microcanonical temperature $T = 1/k_B \beta$ diverges through the transition. Microcanonical deviations from the canonical and grand-canonical predictions are expected around this transition region where the variation of the density of states with energy is minimal.

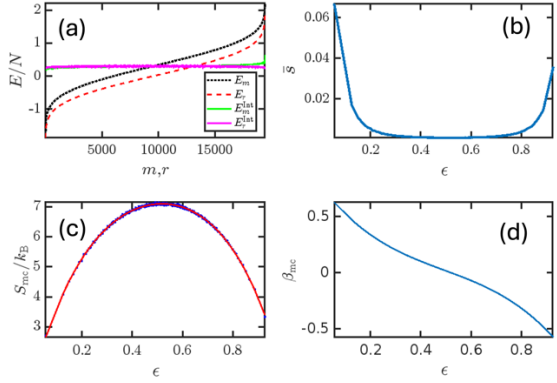

Figure SM1: (a) The interaction free spectrum $\{E_r\}$ (red line), the exact spectrum $\{E_m\}$ (black line), and the interaction energies $E_m^{Int} = \langle m|\hat{H}_{Int}|m\rangle$ (green line) and $E_r^{Int} = \langle r|\hat{H}_{Int}|r\rangle$ (violet line) of an oligomer with $M = 8$, $N = 10$ and $\chi = 3$. (b) The mean spacing between adjacent levels versus the rescaled energy $\epsilon = (E - E_{min})/(E_{max} - E_{min})$. (c) The microcanonical entropy versus internal energy $E_r$; The red line is the best fit to the numerical data (blue points) (d) The inverse temperature versus the total internal energy $E_r$.

*Estimation of nonlinear interaction strength for predominantly ergodic dynamics* – It is instructive to define the average (per mode) optical power density $p = P/M$ and the associated averaged energy density $h = E/M$.

The ergodic nature of the spreading inside the energy shell that stems from the violation of integrable dynamics can be guaranteed by enforcing chaoticity of the underlying classical dynamics generated by Eq. (2) of the main text. The following simple argument can provide a qualitative estimation of the critical nonlinearity strength above which KAM integrability is destroyed leading to a "predominantly" chaotic behavior. The nonlinear frequency shift is $\delta \varepsilon \sim \chi |\psi_l|^2 \sim \chi \cdot p$ (we have used the normalization $1 = \sum_{l=1}^{M} |\psi_l|^2 \sim M \langle |\psi_l|^2 \rangle = M \cdot p \to \langle |\psi_l|^2 \rangle \equiv p = 1/M$). Chaotic dynamics and thermalization occurs when this shift is comparable to a typical energy spacing between frequencies of the linear system i.e. $\delta \varepsilon \geq \Delta \sim \frac{\varepsilon_{max} - \varepsilon_{min}}{M} = \frac{\Delta \varepsilon}{M}$. Substituting the expression for the frequency shift, we have the condition $\Delta \varepsilon \sim \chi \cdot p \geq \Delta \sim \frac{\Delta \varepsilon}{M} \to \frac{\chi p M}{\Delta \varepsilon} \geq 1 \to = \chi \geq \frac{\Delta \varepsilon}{pM}$ which leads to $\chi^* \sim \Delta \varepsilon$.

*Detection of chaotic domains* – Ergodic dynamics, wherein initial preparations uniformly explore the energy shell is a prerequisite for thermalization. We should therefore identify the parameter and energy regimes wherein the system is chaotic. Classically, chaos may be quantified by evaluation of the maximum Lyapunov exponent (mLE) $\lambda_m$ following Ref. [43] of the main text. Given an initial configuration $\{\psi_l(t=0)\}$ and a small variation $\{\delta\psi_l(t=0)\}$ we compute the time evolution of the finite time mLE

$$X_1(t) = \frac{1}{t} \ln\left[\frac{\|\delta\psi(t)\|}{\|\delta\psi(0)\|}\right]; \quad (SM3)$$

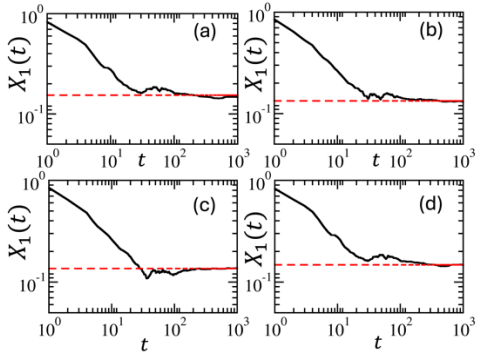

Figure SM2: Finite time maximum Lyapunov exponent $X_1(t)$ versus time for the system of Eq. (2) with the parameters used in Fig. 5 of the main text and internal energy values: (a) $E \approx -0.103$; (b) $E = -0.03$; (c) $E = 0.097$; (d) $E = 0.297$. The red dashed line indicates the asymptotic value $\lambda_m > 0$.

and from there we extract the mLE as $\lambda_m = \lim_{t\to\infty} X_1(t)$. To eliminate fluctuations, we have further average $X_1(t)$ over ten different initial configurations. For regular trajectories $X_1(t)$ tends to zero following a power law $X_1(t) \propto t^{-1}$. For chaotic trajectories $X_1(t)$ converges to a finite mLE $\lambda_m > 0$. In Fig. SM2 we report some typical finite time mLEs for the energies $E$ and for the value of nonlinearity that has been used in Fig. 4 of the main text. In all cases, we were able to confirm that the finite mLE $\lambda_m > 0$ and, therefore, the underlying dynamics is chaotic.

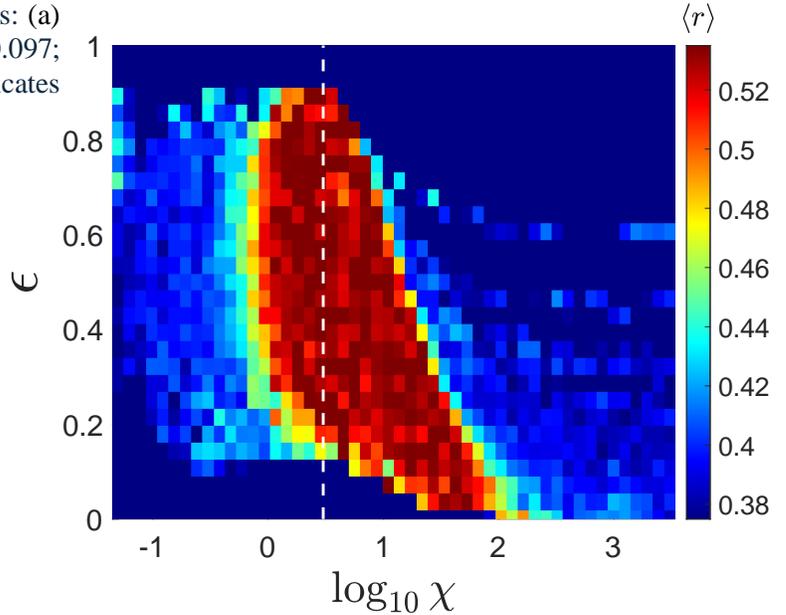

Figure SM3: A density plot of the spacing correlation measure versus the nonlinearity strength $\chi$ and the rescaled energy $\epsilon = (E - E_{min})/(E_{max} - E_{min})$. The white dashed line indicate the $\chi$-value which was used in the calculations shown in Fig. 4 of the main text.

Quantum mechanically, the spectral hallmark of chaos is the Wigner-Dyson level-spacing statistics, as compared to the Poisson statistics in integrable systems. Several measures may be used to quantify the degree of chaoticity, one of them being the spacing correlation measure $\langle r \rangle = \langle \min(s_n, s_{n+1})/\max(s_n, s_{n+1}) \rangle$ where $\{s_n\}$ are adjacent spacings and $\langle . \rangle$ denotes averaging over all spacing in a given energy shell. The value $\langle r \rangle \approx 0.536$ indicates the Gaussian orthogonal ensemble (GOE) random-matrix statistics anticipated for time-reversal-invariant systems exhibiting chaotic dynamics. By contrast, the value $\langle r \rangle \approx 0.386$ indicates integrability. In Fig. SM3 we show the $\langle r \rangle$ measure as a function of the interaction parameter $\chi$ and the rescaled energy $\epsilon = (E - E_{min})/(E_{max} - E_{min})$ where $E_{min}$ and $E_{max}$ are the bottom and top values of energies for each value of $\chi$. For the interaction parameter used in Fig. 4 of the main text (dashed white line) chaos prevails through most shells, except at very low and very high energies. There is an excellent agreement between the chaotic domain and the energy range where thermalization is demonstrated in Fig. 4 of the main text. We have compared the $\langle r \rangle$ measure to other chaos indicators such as the Kulback Leibler divergence of the level spacing distribution from Wigner-Dyson, the Kullback Leibler divergence of the detailed $r$ distribution from the known GOE expectation, and the participation number of interaction-free eigenstates in the exact basis. All these measures give the same chaotic domain.

*Existence of non-conventional thermal state for other bosonic configurations* – We have performed the same analysis as the one reported in the main text for interacting bosonic systems described by the Hamiltonians Eq. (1,2) whose $M$−dimensional connectivity matrix has elements $J_{lm} = J_{ml}$ that are given by a Gaussian distribution with zero mean and a variance $\sigma^2 = 1/M$. Such matrices belong to the so-called Gaussian Orthogonal Ensemble (GOE) and are typically considered prototype mathematical models for the statistical description of complex system with time-reversal symmetry. We have found that the non-conventional occupation statistics, occurs in both quantum and classical systems with GOE connectivity matrix.

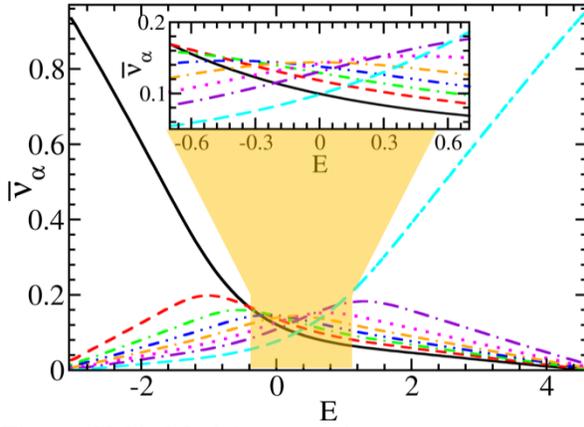

Figure SM4: Modal occupations $\bar{\nu}_\alpha$ versus energy for an oligomer with $M = 8$ and $N = 50$ whose connectivity matrix is taken from GOE. Various lines indicate the various modes: $\bar{\nu}_1$(solid black), $\bar{\nu}_2$ (dashed red), $\bar{\nu}_3$(dashed-dotted green), $\bar{\nu}_4$(dashed-double dotted blue), $\bar{\nu}_5$(doubled dashed- dotted orange), $\bar{\nu}_6$(dotted magenta), $\bar{\nu}_7$(long dashed-dotted violet), $\bar{\nu}_8$(long-dashed light blue). The orange highlighted domain indicates the energy range where scrambling of the modal occupancies occurs. In this domain the microcanonical modal occupancies differ drastically from the BE or the RJ statistics (not shown). Inset: Magnification of the scrambling domain.

In Fig. SM4 we report the microcanonical calculations for such system with $M = 8$ and $N = 50$. The scrambling domain is again evident and occurs for the central section of shell energies.

The quantum calculations of the occupation number $\bar{\nu}_\alpha$ for an oligomer with $M = 8, N = 10$ and with nonlinearity $\chi = 3$ are shown in Fig. SM5 with color filled circles. The connectivity matrix $J$ is taken from the GOE. In these figures we also overplot the microcanonical results (open black diamonds) together with the BE (black solid lines) distribution. The disagreement between the quantum results and the BE is evident. At the same time, the microcanonical results agree nicely with our quantum calculations whenever chaotic dynamics prevails. The degree of chaoticity for each shell is indicated by the color coding and was determined by the evaluation of Shannon entropy (alternative measure to participation number) $S_{r_0} = \left(\sum_m \mathcal{L}_{r_0,m} \ln \mathcal{L}_{r_0,m}\right)/\log(0.48 \cdot D)$ where $\mathcal{L}_{r_0,m}$ is the LDOS of the unperturbed eigenstates in the exact basis.

Finally, in Fig. SM6 we report the mean occupation statistics for a classical system Eq. (2) with an $M-$dimensional connectivity matrix taken from the GOE ensemble. The results of the dynamical simulations (filled black squares) are nicely described by the microcanonical predictions (green diamonds). Instead, the expected RJ statistics (blue dashed line) cannot capture the results of the dynamical simulations.

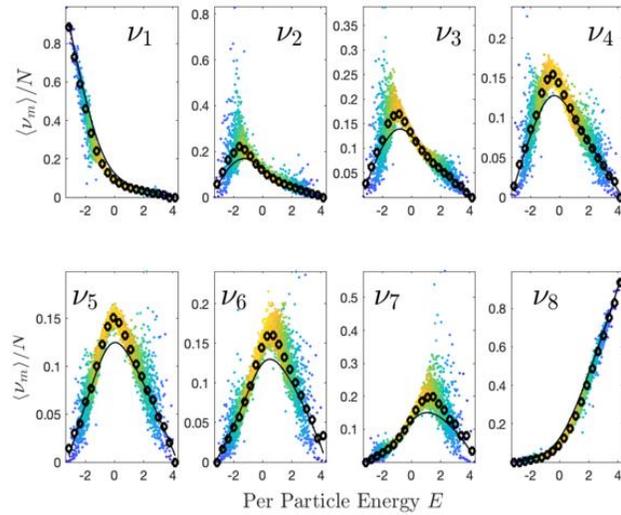

Figure SM5: Modal occupations $\nu_\alpha$ vs. the shell energies $E$ for a fully connected oligomer with $M = 8, N = 10, \chi = 3$. Quantum calculations (colored filled circles), microcanonical results (black diamonds) and BE statistics (black solid line). The color-scale in the quantum results indicates the value of Shannon entropy $S_{r_0}$ of the corresponding eigenmodes. The regimes of highest $S_{r_0}$ (yellow) indicate an ergodic spreading of the unperturbed states while low $S_{r_0}$ values (blue) indicate absence of ergodic spreading. In the former case the system can reach thermalization while in the latter it does not. The deviation of the quantum results from the BE expectations is apparent for $\nu_{3,4,5,6}$ in the energy range $E \approx [-1,1]$. Instead, the quantum results agree nicely with the microcanonical calculations.

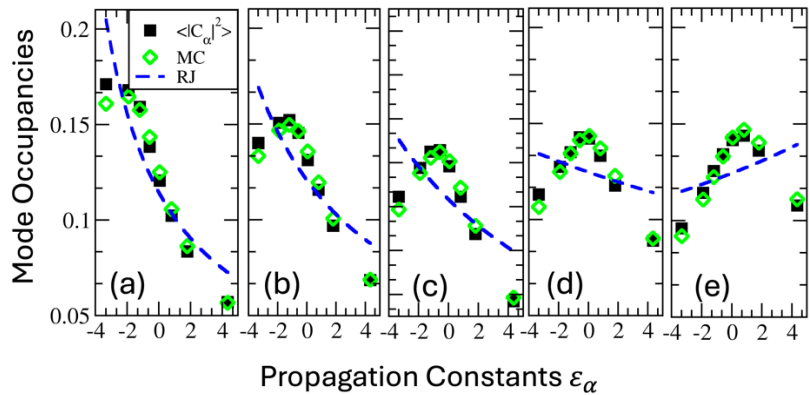

Figure SM6: Modal power occupancies at thermal equilibrium of an initial beam excitation that propagates in a multimode array with $M = 8$ modes and $\chi = 3$. The total initial power is $P = 1$. The values of the internal energy of the initial beam are: (a) $E \approx -636$; (b) $E \approx -0.406$; (c) $E \approx -0.256$; (d) $E \approx -0.103$; (e) $E \approx 0.124$. The results of the dynamical simulations (black square), the microcanonical calculations (green diamonds) and the RJ statistics (blue dashed lines) are indicated in all cases. As the total internal energy increases towards the center of the scrambling domain, the simulations show a stronger deviation from the grand-canonical RJ predictions. Instead, the microcanonical calculations describe nicely the numerical data in all energy regimes.